\documentclass[journal]{IEEEtran}

\ifCLASSINFOpdf
\else
\fi
\usepackage{amsmath, amssymb, bm, cite, epsfig}
\usepackage{epstopdf}
\usepackage{graphicx}
\usepackage{array}
\usepackage{multirow}
\renewcommand{\arraystretch}{1.5}

\graphicspath{{figures/}}

\def\beq{\begin{equation}}
\def\eeq{\end{equation}}
\def\beqa{\begin{eqnarray}}
\def\eeqa{\end{eqnarray}}
\def\beqan{\begin{eqnarray*}}
\def\eeqan{\end{eqnarray*}}

\def\PL{\mathrm{PL}}
\def\dB{\mathrm{dB}}

\def\tm1{t\! - \! 1}
\def\tp1{t\! + \! 1}

\def\PL{\mathrm{PL}}
\def\dB{\mathrm{dB}}
\def\dBm{\mathrm{dBm}}
\def\mW{\mathrm{mW}}
\def\Pt{\mathrm{Pt}}
\def\Pr{\mathrm{Pr}}

\usepackage{tikz}
\usetikzlibrary{calc}

\begin{document}
\title{Synthesizing Omnidirectional Antenna Patterns, Received Power and Path Loss from Directional Antennas for 5G Millimeter-Wave Communications}

\author{\IEEEauthorblockN{Shu Sun, George R. MacCartney Jr., Mathew K. Samimi, and Theodore S. Rappaport\\}
\IEEEauthorblockA{NYU WIRELESS and Polytechnic School of Engineering\\New York University\\
Brooklyn, NY 11201 USA\\
Email: \{ss7152,gmac,mks,tsr\}@nyu.edu}

\thanks{Sponsorship for this work was provided by the NYU WIRELESS Industrial Affiliates program and NSF research grants 1320472, 1302336, and 1555332.}
}
\maketitle
\begin{tikzpicture}[remember picture, overlay]
\node at ($(current page.north) + (0,-0.25in)$) {S. Sun, G. R. MacCartney, Jr., M. K. Samimi, and T. S. Rappaport, \rq\rq{}Synthesizing Omnidirectional Antenna Patterns, Received Power};
\node at ($(current page.north) + (0,-0.4in)$) {and Path Loss from Directional Antennas for 5G Millimeter-Wave Communications,\rq\rq{} to appear in \textit{2015 IEEE Global}};
\node at ($(current page.north) + (0,-0.55in)$) {\textit{Communications Conference (Globecom)}, Dec. 2015.};
\end{tikzpicture}

\begin{abstract}
Omnidirectional path loss models are vital for radio-system design in wireless communication systems, as they allow engineers to perform network simulations for systems with arbitrary antenna patterns. At millimeter-wave frequencies, channel measurements are frequently conducted using steerable high-gain directional antennas at both the transmitter and receiver to make up for the significant increase in free space path loss at these frequencies compared to traditional cellular systems that operate at lower frequencies. The omnidirectional antenna pattern, and resulting omnidirectional received power must therefore be synthesized from many unique pointing angles, where the transmit and receive antennas are rotated over many different azimuth and elevation planes. In this paper, the equivalent omnidirectional antenna pattern and omnidirectional received power are synthesized by summing the received powers from all measured unique pointing angles obtained at antenna half-power beamwidth step increments in the azimuth and elevation planes, and this method is validated by demonstrating that the synthesized omnidirectional received power and path loss are independent of antenna beamwidth, through theoretical analyses and millimeter-wave propagation measurements using antennas with different beamwidths. The method in this paper is shown to provide accurate results while enhancing the measurement range substantially through the use of directional antennas.
\end{abstract}
\IEEEpeerreviewmaketitle

\section{Introduction}
Millimeter-wave (mmWave) is envisioned as a promising frequency band for fifth generation (5G) mobile and backhaul broadband wireless communications, where a massive amount of raw bandwidth will enable multi-gigabit-per-second data rates\cite{WillWork:TSR13, Pi11, Rap15,Rap15:TCOM}. The significant increase in free space path loss at mmWave frequencies  is conveniently overcome by using high-gain directional antennas at the base station and/or mobile handset~\cite{Rap14}, which provide sufficient gain to complete mmWave links over 500 m or so, as shown in~\cite{WillWork:TSR13,MacCartney14:1,Rap15:TCOM}. MmWave propagation measurements are needed to accurately characterize channels and create statistical channel models, necessary for proper design of wireless systems and protocols.  While electrically-steerable adaptive antennas will be used in 5G mmWave systems~\cite{WillWork:TSR13, Gutierrez09,Zhou14, Roh14}, such antennas are not yet conveniently available for researchers at most mmWave frequencies. In the meantime, many researchers are using mechanically rotatable horn antennas to measure the channel at a wide range of mmWave frequencies \cite{WillWork:TSR13, MacCartney14:1, Hur14, Raj12}. In lieu of directional antennas, omnidirectional antennas are feasible and an alternative to study propagation channels, but these require much greater transmit power to achieve the same measurement range (or maximum measurable path loss). 

To synthesize what an omnidirectional receiver antenna would receive, and to generate omnidirectional path loss models where arbitrary directional antenna patterns can be implemented for specific measurement or simulation applications, an accurate procedure is required to recover omnidirectional path loss from directional path loss measurements, where the directional transmit (TX) and receive (RX) antennas are typically rotated over many azimuth and elevation planes to emulate omnidirectional antennas. The omnidirectional antenna pattern was synthesized to provide omnidirectional path loss models at 28 GHz and 73 GHz using directional measurements\cite{MacCartney14:2} by summing up the received powers (in milli-Watts) at each and every unique measured non-overlapping TX and RX antenna pointing angle combination. The omnidirectional path loss was calculated by subtracting the summed received power from the transmit power, with the TX and RX antenna gains removed from each directional measurement (to provide the equivalent 0 dB isotropic gain for each directional measurement power contribution):
\begin{equation}\label{IREq}\begin{split}
&\PL_{i,j}[\dB]= \Pt_{i,j}[\dBm]-\\
&10\log_{10}\left( \sum_n \sum_m \sum_l \sum_k \Pr_{i,j}(\phi_k, \theta_l, \varphi_m,\vartheta_n)[\mW]\right)
\end{split}\end{equation}
where $\PL_{i,j}$, $\Pt_{i,j}$, and $\Pr_{i,j}$ denote the synthesized omnidirectional path loss, synthesized omnidirectional transmit power, and directional received power for a specific and unique angle of departure (AOD) and angle of arrival (AOA) combination (with the antenna gains removed) from the $i\textsuperscript{th}$ TX to the $j\textsuperscript{th}$ RX, respectively. $\phi,\theta,\varphi,\vartheta$ are the azimuth AOA, elevation AOA, azimuth AOD, and elevation AOD, respectively.
This paper validates the method used in \cite{MacCartney14:2} for synthesizing the omnidirectional antenna pattern (while reaping the benefit of directional gain) and recovering omnidirectional received power and path loss. 

Path loss models are essential for estimating the attenuation of propagating signals over distance, in order to adequately design wireless communication systems. For new wireless networks, it is desirable to obtain as much measurement range as possible, to ensure accuracy of the models out to large distances. Two common empirical path loss models are the close-in (CI) free space reference distance model and the floating-intercept (FI) (\textit{alpha-beta}) model~\cite{Rap15:TCOM}. 

\subsubsection{Close-in Free Space Reference Distance Model}
The CI model considers free space propagation ($n=2$) up to a close-in anchor distance $d_0$, beyond which a minimum mean-square error (MMSE) fit is performed on the measured path losses, resulting in the path loss exponent (PLE) $n$ that quantifies the rate of signal attenuation over log-distance. The $d_0=1$ m CI model is given by Eq.~\eqref{eq:FSPL}:
\begin{align}\label{eq:FSPL}
&\PL(d)[\dB] =\PL(d_0)+10n\log_{10}\left(\frac{d}{d_0}\right)+X_{\sigma}\\
&\PL(d_0)[\dB] = 20\log_{10}\left(\frac{4\pi d_0}{\lambda}\right)
\end{align}

\noindent where $X_{\sigma}$ is a zero-mean Gaussian distributed random variable (in dB) with standard deviation $\sigma$ (also in dB), that describes the random large-scale shadowing or variations in signal level around the mean path loss at any distance $d$ from the MMSE fit~\cite{Rappaport:Wireless2nd,Rap15:TCOM}. 

The CI model employs a 1 m physical free space reference distance and was shown in\cite{Rap15:TCOM} to be much more stable for extrapolating the model outside of the measured distance range. The single path loss parameter ($n$, the PLE) provides insight into physical propagation characteristics. Previous UHF (Ultra-High Frequency)/Microwave models used a close-in reference distance $d_0$ of 1 km or 100 m since base station towers were tall without any nearby obstructions and inter-site distances were on the order of many kilometers for those frequency bands\cite{Hata:TVT80,And95}. However, $d_0$ = 1 m is used in mmWave path loss models since base stations will be shorter or mounted indoors, and closer to obstructions~\cite{WillWork:TSR13,Rap15:TCOM}, and signal propagation can be regarded as free-space propagation within 1 m of the transmitter. Further, the 1 m CI model enables a convention that is easily used for allowing 10$n$ to describe path loss in terms of decades of distances beginning at 1 m, and provides for an international reference\cite{Rap15:TCOM}.

\subsubsection{\textit{Floating-Intercept} Model}

The FI model is used by researchers who helped develop WINNER II and 3GPP channel models~\cite{WinnerII, 3GPP}. The FI model employs the best fit line to the measured data using a least-squares regression method where two parameters, an intercept $\alpha$ (dB) and slope $\beta$, are extracted to form the equation~\cite{Rap15:TCOM,PL5G:GMAC13}:
\begin{equation}\label{eq:AB}
\PL[\dB] = \alpha+10\cdot\beta\log_{10}(d)+X_{\sigma}
\end{equation}
where $X_{\sigma}$ is a zero-mean Gaussian distributed random variable (in dB) with standard deviation $\sigma$ (also in dB). When compared with the 1 m CI model, the FI model usually (but not always) reduces the shadowing standard deviation by a fraction of a dB~\cite{Rap15:TCOM}, little difference compared to the typical value of 7 dB to 12 dB for $X_{\sigma}$ in non-line-of-sight (NLOS) environments\cite{Rap15:TCOM}, thus suggesting the CI model is sufficient and easier to use, while avoiding the extra complexity of a second model parameter. Moreover, the FI model lacks a physical anchor to the transmitted power (or true path loss) in the first meter of transmission, and is only valid over the measured distances. Additionally, the slope $\beta$ is not the same as a PLE, rather it is a slope of the best fit line to a measured set of data. In some cases, the measured data are so sparse or clustered in distance such that the slope $\beta$ can be very close to zero or even negative, and the intercept $\alpha$ may be much higher than theoretical free space path loss at short distances\cite{Rap15:TCOM,PL5G:GMAC13, Sulyman:ComMag14}. 

The 1 m CI model is more robust for data sets as it considers only one path loss parameter $n$, as opposed to two parameters $(\alpha,\beta)$ in the FI model, and allows comparisons of results across different distances, environments, and research groups in a global manner. Note that the 1 m CI model makes physical sense even in NLOS environments, because the first few meters from the transmitter will still be line-of-sight (LOS) (considering typical environments where the radio waves will propagate a number of meters before encountering the first obstacle)\cite{Rap15:TCOM}. Further, it has been recently shown that there is virtually no difference between using the FI model and the 1 m CI model in NLOS environments when considering a probabilistic-type path loss model taking into account the LOS and NLOS probabilities, showing the value of a universal 1 m CI model for all distances and environments at mmWave frequencies\cite{MKS:WCL15}. 

\section{Measurement Procedure}
In the summers of 2012 and 2013, two outdoor propagation measurement campaigns were conducted at 28 GHz and 73 GHz, respectively, in downtown Manhattan, New York, using similar 400 Megachips-per-second (Mcps) spread spectrum sliding correlator channel sounders and directional steerable horn antennas at both the TX and RX to investigate mmWave channel characteristics in a dense urban micro-cell (UMi) environment \cite{WillWork:TSR13, MacCartney14:1}. The measurements provided an RF null-to-null bandwidth of 800 MHz and multipath time resolution of 2.5 ns, which guarantees that a majority of multipath components (MPCs) can be distinguished. In the 28 GHz measurements, three TX locations (heights of 7 m and 17 m) and 27 RX locations (heights of 1.5 m) were selected to conduct the measurements\cite{Rap15:TCOM}. Two types of horn antennas were employed: a 24.5 dBi-gain narrowbeam horn antenna with 10.9$^{\circ}$ and 8.6$^{\circ}$ half-power beamwidths (HPBWs) in the azimuth and elevation planes, respectively, and a 15 dBi-gain widebeam horn antenna with 28.8$^{\circ}$ and 30$^{\circ}$ HPBWs in the azimuth and elevation planes, respectively. The narrowbeam antenna was always utilized at the TX locations, and five of the RX locations used both the narrowbeam and widebeam antennas, including two LOS and three NLOS locations. For nine out of the ten measurement sweeps for each TX-RX location combination (except the two LOS RX locations), the RX antenna was sequentially swept over the entire azimuth plane in increments of one HPBW at elevation angles of 0$^{\circ}$ and $\pm20^{\circ}$ about the horizon, so as to measure contiguous angular snapshots of the channel impulse response over the entire 360$^{\circ}$ azimuth plane, while the TX antenna remained at a fixed azimuth and elevation angle. The TX antenna was swept over the azimuth plane in the last measurement sweep~\cite{WillWork:TSR13}. 

In the 73 GHz measurements, five TX locations (heights of 7 m and 17 m) and 27 RX locations were used, with RX antenna heights of 2 m (mobile scenario) and 4.06 m (backhaul scenario), yielding a total of 36 TX-RX location combinations for the mobile (access) scenario and 38 combinations for the backhaul scenario. A pair of 27 dBi-gain rotatable directional horn antennas with a HPBW of 7$^{\circ}$ in both azimuth and elevation planes was employed at the TX and RX. For each TX-RX location combination, TX and RX antenna azimuth sweeps were performed in steps of 8$^{\circ}$ or 10$^{\circ}$ at various elevation angles. In the measurement system, the total time of acquiring a power delay profile (PDP) (including recording and averaging 20 instantaneous PDPs) was 40.94 ms$\times$20 = 818.8 ms, where 40.94 ms was the time it took to record a single instantaneous PDP (The times were dilated based on the sliding correlation method)\cite{Rap15:TCOM}. Additional measurement procedures, channel modeling results, and hardware specifications can be found in~\cite{WillWork:TSR13,MacCartney14:1,Rap15:TCOM, Rap15}.

\section{Power Synthesizing Theory}
The method for synthesizing the omnidirectional receive antenna pattern, and thus the omnidirectional received power introduced in Section I is now theoretically validated step-by-step. First, assuming omnidirectional antennas are used at both the TX and RX, if $N$ MPCs arrive at the RX, then the received signal $r(t)$ can be expressed as\cite{Rappaport:Wireless2nd}:
\begin{equation}\label{eq:omniSig}
r(t)=\sum\limits_{n=1}^N a_{n}e^{j\Phi_{n}}\delta (t-\tau_{n})
\end{equation}

\noindent where $a_{n}$, $\Phi_{n}$, and $\tau_{n}$ are the amplitude, phase, and propagation time delay of the $n\textsuperscript{th}$ MPC, respectively. $\delta(\cdot)$ denotes the Kronecker delta function. The received power is:
\begin{equation}\label{eq:omniP1}
\begin{split}
P_{tot}&=\sum\limits_{t=\tau_1}^{\tau_N} |r(t)|^2 \\
&=\sum\limits_{t=\tau_1}^{\tau_N}\sum\limits_{i=1}^N \sum\limits_{k=1}^N a_{i}a_{k}e^{j(\Phi_{i}-\Phi_{k})}\delta (t-\tau_{i})\delta (t-\tau_{k})\\
&=\sum\limits_{i=1}^N\sum\limits_{k=1}^Na_{i}a_{k}e^{j(\Phi_{i}-\Phi_{k})}\delta (\tau_{i}-\tau_{i})\delta (\tau_{i}-\tau_{k})\\
&=\sum\limits_{i=1}^N\sum\limits_{k=1}^Na_{i}a_{k}e^{j(\Phi_{i}-\Phi_{k})}\delta (\tau_{i}-\tau_{k})
\end{split}
\end{equation}

\noindent The time delay $\tau$ differs for each resolvable MPC, hence the double sum in \eqref{eq:omniP1} is zero for $k\neq i$. Thus, the received power can be simplified to:
\begin{equation}\label{eq:omniP2}
P_{tot}=\sum\limits_{i=1}^N {a_i}^2e^{j(\Phi_{i}-\Phi_{i})}=\sum\limits_{i=1}^N {a_i}^2
\end{equation}

\noindent Next, suppose directional antennas are used at the same TX and RX locations with antenna gains $G_{T}$ and $G_{R}$ in linear units, respectively. For one AOD and AOA pointing angle combination, a subset of MPCs shown in \eqref{eq:omniSig} will arrive at the RX. Assuming there are totally $Q$ AOD and AOA combinations over the $4\pi$ steradian sphere for a TX-RX pair, and $M_q$ MPCs reach the RX for the $q^{th}$ AOD and AOA combination, where the value of $M_q$ is dependent on the AOD and AOA combination and $\sum\limits_{q=1}^Q M_q=N$, then the received signal for the $q^{th}$ AOD and AOA combination is:
\begin{equation}\label{eq:directSig}
r_{q}(t)=\sqrt{G_{T}}\sum\limits_{m=1}^{M_q}\sqrt{G_{R}}\cdot a_{m}e^{j\Phi_{m}}\delta (t-\tau_{m})
\end{equation}

\noindent Note that each MPC in \eqref{eq:directSig} corresponds to an MPC in \eqref{eq:omniSig}. Further note the value of this method, where the square root of the directional antenna power gains are included in the received voltage signal for each directional measurement, offering inherent gain and distance extension not available with a typical omnidirectional antenna having 0 dBi gain. Using the same approach derived in \eqref{eq:omniP2} from \eqref{eq:omniSig}, the received power for the $q^{th}$ AOD and AOA combination is:
\begin{equation}\label{eq:directP1}
P_{q}=G_{T}G_{R}\sum\limits_{l=1}^{M_q} {a_l}^2
\end{equation}

\noindent By performing an exhausive antenna sweep over all $Q$ possible AOD and AOA combinations without spatial overlap, i.e., individual measurements are separated by one HPBW in both the azimuth and elevation planes, the sum of received power over all unique AOD and AOA combinations yields:
\begin{equation}\label{eq:directP2}
\sum\limits_{q=1}^Q P_{q}=G_{T}G_{R}\sum\limits_{l=1}^N {a_l}^2
\end{equation}
which is equivalent to the omnidirectional received power in \eqref{eq:omniP2} after removing the antenna gains. Therefore, the sum of received powers from non-overlapping angles in the azimuth and elevation planes from directional antenna measurements results in the omnidirectional received power, after removing antenna gains. The antenna gains offer significant range extension and much greater path loss measurement range than would standard omnidirectional antennas, thus demonstrating the advantage of using directional antennas to synthesize omnidirectional antenna patterns.

The method for synthesizing omnidirectional received power can also be validated by considering antenna radiation patterns. The far-field power radiation pattern of a horn antenna can be approximated by \cite{www}:
\begin{equation}\label{eq:AntPattern}
\begin{split}
f(\phi, \theta)=G\left[\rm{sinc}^2\left(a\cdot\sin(\phi)\right)\cos^2(\phi)\right]\\
\cdot\left[\rm{sinc}^2\left(b\cdot\sin(\theta)\right)\cos^2(\theta)\right]
\end{split}
\end{equation}

\noindent where $\phi$ and $\theta$ represent the azimuth and elevation angles with respect to (w.r.t.) the antenna boresight, respectively, $f(\phi, \theta)$ denotes the radiation power density at the azimuth angle $\phi$ and elevation angle $\theta$, $G$ represents the boresight gain of the antenna, and $a$ and $b$ are functions of the azimuth (AZ) and elevation (EL) HPBWs of the horn antenna, respectively, i.e., 
\begin{equation}\label{eq:antPat1}
\begin{split}
\rm{sinc}^2 \Bigg( a\cdot\sin \Big(\frac{\rm{HPBW}_{AZ}}{2}\Big) \Bigg)\cos^2\left(\frac{\rm{HPBW}_{AZ}}{2}\right)=\frac{1}{2}
\end{split}
\end{equation} 
\begin{equation}\label{eq:antPat2}
\begin{split}
\rm{sinc}^2 \Bigg( b\cdot\sin \Big(\frac{\rm{HPBW}_{EL}}{2}\Big) \Bigg)\cos^2\left(\frac{\rm{HPBW}_{EL}}{2}\right)=\frac{1}{2}
\end{split}
\end{equation}

\begin{figure}
\centering
\includegraphics[width = 90mm]{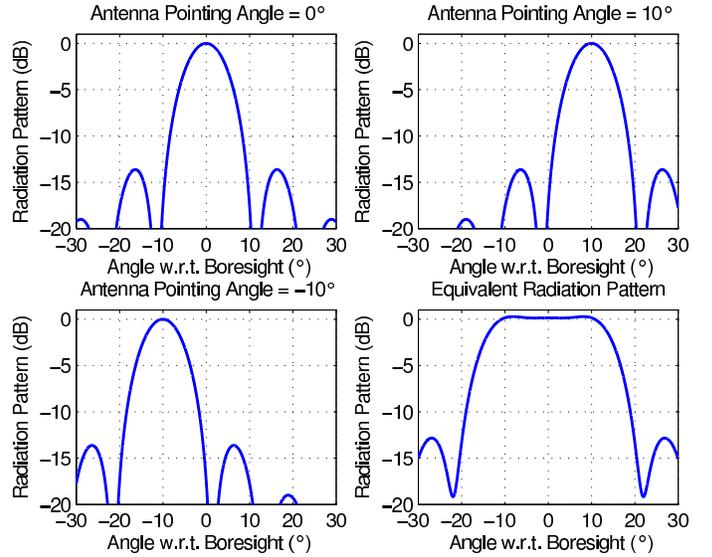}
\caption{Normalized antenna pattern in the azimuth plane for a horn antenna with an azimuth HPBW of 10$^{\circ}$ at azimuth pointing angles of 0$^{\circ}$, 10$^{\circ}$, and -10$^{\circ}$ with respect to the boresight angle, and the normalized equivalent radiation pattern by linearly adding the antenna power amplitude patterns at these three adjacent angles.}
\label{fig_omp}
\end{figure}

\begin{figure}
\centering
\includegraphics[width = 90mm]{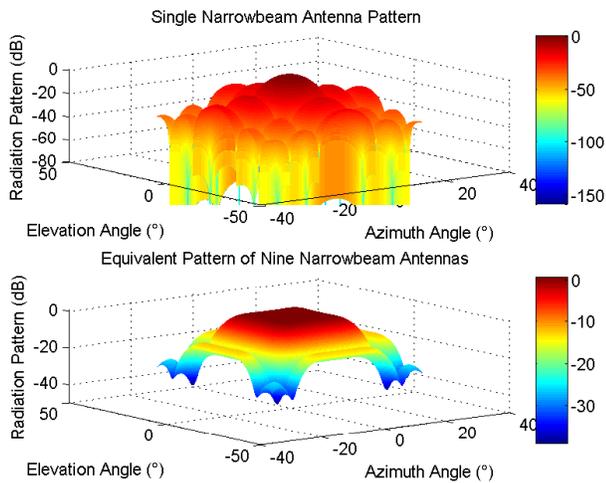}
\caption{Normalized antenna pattern in both the azimuth and elevation planes for a horn antenna with an azimuth HPBW of 10$^{\circ}$ and an elevation HPBW of 8$^{\circ}$, pointing at the boresight angle, and the normalized equivalent pattern by overlapping the patterns at all the angle combinations of 0$^{\circ}$, 10$^{\circ}$, and -10$^{\circ}$ in the azimuth plane and 0$^{\circ}$, 8$^{\circ}$, and -8$^{\circ}$ in the elevation plane.}
\label{fig_omp3d}
\end{figure}

\noindent For instance, if the azimuth HPBW of a horn antenna is 10$^{\circ}$ (i.e., 0.17 radians), then $a$ = 5.06. Fig.~\ref{fig_omp} displays the normalized antenna azimuth power pattern for a horn antenna with an azimuth HPBW of 10$^{\circ}$ at an elevation angle of 0$^{\circ}$ and at azimuth angles of 0$^{\circ}$, 10$^{\circ}$, and -10$^{\circ}$ w.r.t. the boresight angle, and the normalized equivalent widebeam antenna power pattern by linearly combining the power amplitude of each adjacent narrowbeam antenna pattern at the three adjacent azimuth angles. It is clear from Fig.~\ref{fig_omp} that the maximum gain in the normalized equivalent pattern remains constant over the range of -10$^{\circ}$ to 10$^{\circ}$, indicating that if linearly combining the power patterns at antenna pointing angles over the entire azimuth plane, i.e., from 0$^{\circ}$ to 360$^{\circ}$, the resultant antenna gain will become constant over the entire azimuth plane and approximately equal to the boresight gain of the directional horn antenna (actually about 0.25 dB larger than the boresight gain of a single antenna, yet this constant offset is quite small and can be removed), and the same outcome holds for the elevation plane. The normalized three-dimensional (3D) pattern of a single directional antenna and the linearly-combined power pattern using 3$\times$3 directional antennas spaced in consecutive HPBW intervals are illustrated in Fig.~\ref{fig_omp3d}, where the antenna is assumed to have an azimuth HPBW of 10$^{\circ}$ and an elevation HPBW of 8$^{\circ}$, with the normalized equivalent pattern obtained by linearly adding the power patterns at all the angle combinations of 0$^{\circ}$, 10$^{\circ}$, and -10$^{\circ}$ in the azimuth plane and 0$^{\circ}$, 8$^{\circ}$, and -8$^{\circ}$ in the elevation plane. It is evident that in the single antenna pattern, the maximum gain is concentrated only on the boresight angle, while in the synthesized pattern the maximum gain (about 0.5 dB larger than the boresight gain of a single antenna, and again this constant offset is quite small and can be removed) remains constant over the entire 24$^{\circ}$$\times$18$^{\circ}$ angular region. Although the antenna patterns in the simulations shown in Figs.~\ref{fig_omp} and~\ref{fig_omp3d} are canonical, they are nearly identical to the actual measured patterns of the antennas used during the measurements. This same concept can be extracted over the entire $4\pi$ steradian sphere to form an omnidirectional antenna pattern. Therefore, the synthesized directional antenna pattern over the entire $4\pi$ steradian sphere after removing the directional antenna gain will approximate the pattern of an omnidirectional antenna, indicating that it is appropriate to acquire the omnidirectional received power by summing up the powers from directional antennas in all possible non-overlapping directions, with antenna gains removed as in~\eqref{IREq}. Note that some MPCs might be counted twice near the directional antenna pattern edge by adjacent antennas, where the antenna gain is smaller than the boresight gain, when summing the powers from adjacent directional antennas. Nevertheless, by linearly adding the antenna power patterns from -3~$\times$ HPBW to 3~$\times$ HPBW in steps of 0.01$^{\circ}$, as was done in Figs.~\ref{fig_omp} and~\ref{fig_omp3d}, the resulting synthesized antenna pattern becomes reasonably flat, as indicated by the last subfigures in Figs.~\ref{fig_omp} and~\ref{fig_omp3d}, such that the synthesized antenna pattern approximates the pattern of an omnidirectional antenna, where the superposition of adjacent directional antenna patterns ensures MPCs received near the directional antenna pattern edge will experience the same antenna gain as those received at boresight when the omnidirectional pattern is synthesized. 

\section{Synthesizing Procedure and Empirical Results}
\subsection{28 GHz Measurement Data}
In this section, the method for synthesizing the omnidirectional antenna pattern and omnidirectional received power is validated by comparing the measured power obtained from two different directional horn antennas using the 28 GHz measurements. Here, it is shown that the corresponding synthesized powers are independent of antenna beamwidth. A pointing angle measured with the 30$^{\circ}$ HPBW widebeam antenna was selected, and then the measured angle was discretized into nine smaller 10$^{\circ}$ angles in azimuth and elevation planes. Then the power measured with the widebeam antenna was compared with the sum of the powers obtained at the nine smaller angles using the narrowbeam antenna.
Note that the 28 GHz measurements used fixed elevation angles of 0$^{\circ}$, and $\pm20^{\circ}$ as opposed to $\pm10^{\circ}$. Therefore, the unavailable $\pm10^{\circ}$ elevation angles were substituted with the available $\pm20^{\circ}$ elevation angles. As shown in Table~\ref{eleAngle}, and based on 73 GHz measurements in Section~\ref{73 GHz Measurement Data} that measured adjacent antenna patterns, most received power is received at the strongest elevation angle, such that the lack of the immediate adjacent measured antenna pattern does not create a substantial error.

Three NLOS RX locations, i.e., RX 14, RX 16, and RX 19 from measurements in New York City~\cite{WillWork:TSR13} were selected to compare the received powers obtained from widebeam and narrowbeam antennas. The measured data sets were chosen such that the TX antennas were pointed in the same azimuth and elevation pointing directions for both narrowbeam and widebeam antenna azimuth sweeps. The RX elevation angle for the azimuth sweep using the widebeam antenna was 0$^{\circ}$, and the elevations for the narrowbeam antenna sweeps were 0$^{\circ}$, -20$^{\circ}$, and 20$^{\circ}$. Ideally, adjacent azimuth planes should be separated by $\pm$ one elevation HPBW, but due to lack of measured data at the $\pm10^{\circ}$ elevation angles, the $\pm20^{\circ}$ elevation angles were examined instead. Table I details the parameters of the selected measurements, where the azimuth angles are w.r.t. the true north bearing direction, and the elevation angles are w.r.t. the horizon where positive angles denote angles above the horizon. 

For fair comparison, 9.5 dB was added to the received power found using the widebeam antenna as shown in Table~\ref{tbl}, which is the difference in the boresight gain of the narrowbeam and widebeam antennas, to compensate for the smaller antenna gain of the widebeam antenna, so that the only difference between the two types of antennas is the antenna beamwidth. By using Eq.~\eqref{eq:AntPattern} and integrating from -3~$\times$ HPBW to 3~$\times$ HPBW in both azimuth and elevation planes, one would expect the received power using a single widebeam (28.8$^{\circ}$/30$^{\circ}$ azimuth/elevation HPBW) antenna would be around 8.8 times (i.e., 9.4 dB greater than) that of a single narrowbeam (10.9$^{\circ}$/8.6$^{\circ}$ azimuth/elevation HPBW) antenna given the same boresight gain. Moreover, again by using Eq.~\eqref{eq:AntPattern} and integrating from -3~$\times$ HPBW to 3~$\times$ HPBW in both azimuth and elevation planes, and linearly combining the power amplitudes of 3$\times$3 narrowbeam antennas spaced in consecutive HPBW intervals, one can obtain that the power difference between using 3$\times$3 narrowbeam antennas spaced in consecutive HPBW intervals and a single widebeam antenna is only 0.08 dB. 


As shown in Table~\ref{tbl}, the effective received power after summing up the received powers from 3$\times$3 narrowbeam antennas is approximated well with the power obtained using one widebeam antenna. For example, Table~\ref{tbl} shows that when the widebeam antenna is pointed at an azimuth angle of 242$^{\circ}$ at RX 19, the effective received power obtained by computing the area under the PDP using the widebeam antenna is -68.7 dBm, while the effective synthesized received power from the linear power combination of the 3$\times$3 narrowbeam angles is -67.0 dBm, only 1.7 dB higher. Furthermore, the effective received powers over the entire azimuth plane using the widebeam and narrowbeam antennas match well at each of the three RX locations where this test was made, as expected from the theoretical analyses in Section III, with a maximum difference of 2.9 dB, as shown by the last three rows in Table~\ref{tbl}.

\begin{table*}
\renewcommand{\arraystretch}{1.2}
\caption{Measurement parameters and comparison of received power and path loss using narrowbeam and widebeam antennas at the receiver. The TX location is the Kaufman (KAU) building on NYU's Manhattan campus, the AOD elevation is -10$^{\circ}$, and the AOA elevation is 0$^{\circ}$ for the widebeam antenna, while 0$^{\circ}$ and $\pm20^{\circ}$ elevation for the narrowbeam antenna. $W$ and $N$ denote widebeam and narrowbeam antennas, respectively. $\Delta P_{r}$ is the difference in received power using widebeam and narrowbeam antennas.}
\label{table1}
\centering
\begin{tabular}{|c|c|c|c|c|c|c|c|c|c|}
\hline
 \raisebox{-1.5ex}{\textbf{RX ID}} & \raisebox{-1.5ex}{\textbf{T-R Separation (m)}} & \raisebox{-1.5ex}{\textbf{AOD Azimuth ($^{\circ}$)}}  & \multicolumn{2}{c|}{\textbf{AOA Azimuth ($^{\circ}$)}}  
& \multicolumn{2}{c|}{\textbf{Effective $\Pr$ (dBm)}} & \multicolumn{2}{c|}{\textbf{Effective $\PL$ (dB)}} 
& \raisebox{-1.5ex}{\textbf{$\Delta \Pr$ (dB)}}\\\cline{4-9}
&  & & \raisebox{0.0ex}{W} & \raisebox{0.0ex}{N} & \raisebox{0.0ex}{W} &  \raisebox{0.0ex}{N} &  \raisebox{0.0ex}{W} &  \raisebox{0.0ex}{N} & \\
\hline
\raisebox{-1.5ex}{RX 14} & \raisebox{-1.5ex}{82} & \raisebox{-1.5ex}{140}  & 32 & 22, 32, 42  & -55.9 & -62.0 & 135.0 & 141.1 & 6.1\\\cline{4-10}
 & & & 62 & 52, 62, 72 & -60.9 & -56.1 & 140.0 & 135.2 & -4.8\\
\hline
 RX 16 & 97 & 140 & 92 & 82, 92, 102 & -62.9 & -65.1 & 142.0 & 144.2 & 2.2\\
\hline
\raisebox{-1.5ex}{RX 19} & \raisebox{-1.5ex}{175} & \raisebox{-1.5ex}{175}  & 212 & 202, 212, 222  & -63.9 & -63.4 & 143.0 & 142.5 & -0.5\\\cline{4-10}
 & & & 242 & 232, 242, 252 & -68.7 & -67.0 & 147.8 & 146.1 & -1.7\\
\hline
 RX 14 & 82 & 140 & \multicolumn{2}{c|}{All Azimuth Angles} & -54.7 & -52.3 & 133.8 & 131.4 & -2.4\\
\hline
 RX 16 & 97 & 140 & \multicolumn{2}{c|}{All Azimuth Angles} & -61.3 & -64.2 & 140.4 & 143.3 & 2.9\\
\hline
RX 19 & 175 & 175 & \multicolumn{2}{c|}{All Azimuth Angles} & -61.7 & -61.3 & 140.8 & 140.4 & -0.4\\
\hline
\end{tabular}
\label{tbl}
\end{table*}

\begin{figure}
\centering
\includegraphics[width = 90mm]{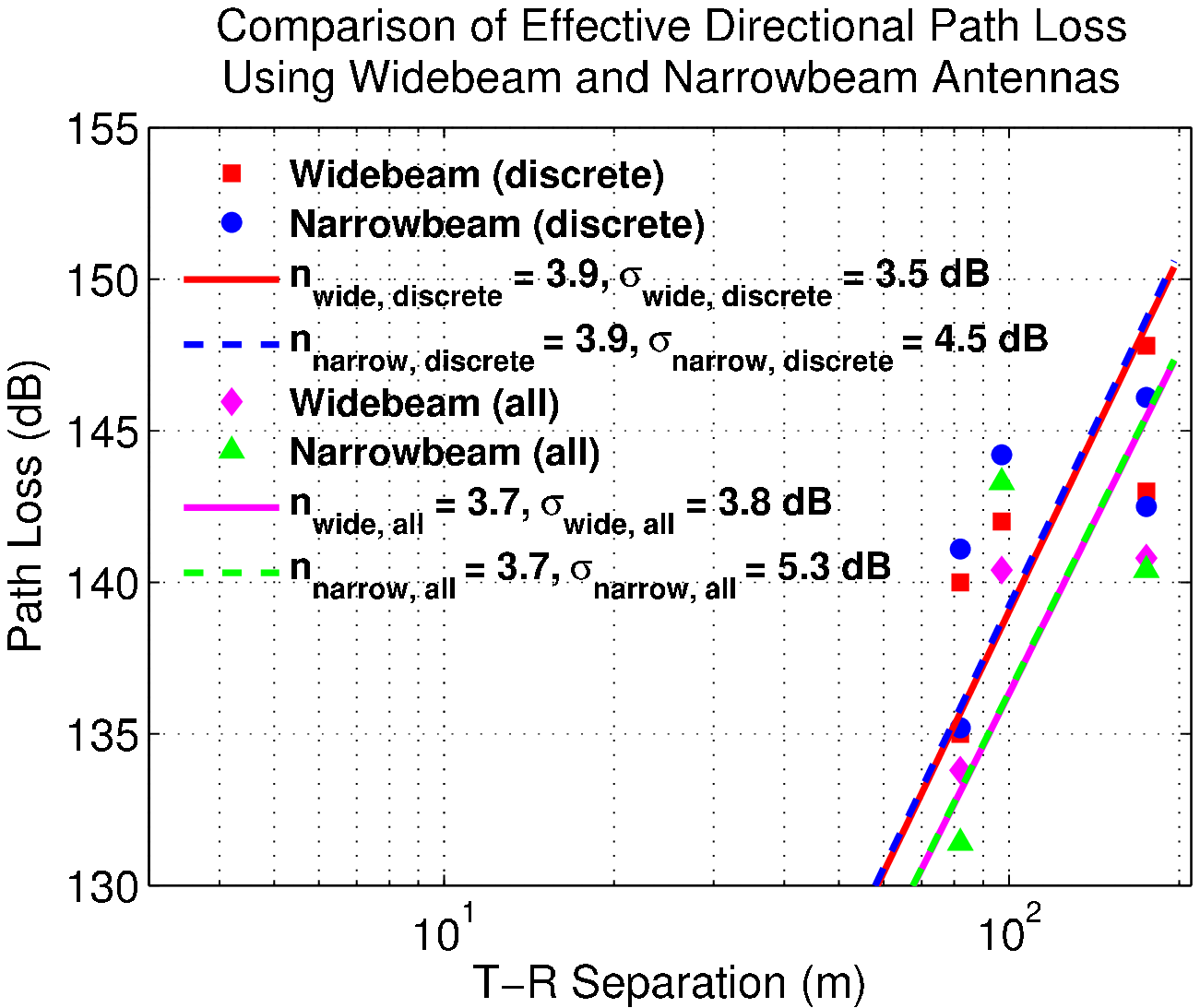}
\caption{Comparison of 28 GHz NLOS effective directional path loss using the widebeam (28.8$^{\circ}$/30$^{\circ}$ azimuth/elevation HPBW) and narrowbeam (10.9$^{\circ}$/8.6$^{\circ}$ azimuth/elevation HPBW) antennas at three RX locations. The \lq\lq{}discrete\rq\rq{} widebeam path loss is obtained using a single widebeam antenna, and the \lq\lq{}discrete\rq\rq{} narrowbeam path loss is synthesized from nine narrowbeam antennas with adjacent antennas separated by $10^{\circ}$ and $20^{\circ}$ in the azimuth and elevation planes, respectively. The \lq\lq{}all\rq\rq{} path loss corresponds to the effective path loss over the entire azimuth plane(s).}
\label{fig_pl}
\end{figure}

Fig.~\ref{fig_pl} is a scatter plot of the 28 GHz effective directional path loss using the widebeam and narrowbeam antennas at the three RX locations, using the 1 m CI path loss model.  The \lq\lq{}discrete\rq\rq{} widebeam path loss is obtained using a single widebeam antenna, and the \lq\lq{}discrete\rq\rq{} narrowbeam path loss is synthesized from 3$\times$3 narrowbeam antennas. The \lq\lq{}all\rq\rq{} widebeam path loss corresponds to the effective path loss over the entire azimuth plane at a 0$^{\circ}$ elevation angle, while the \lq\lq{}all\rq\rq{} narrowbeam path loss is synthesized from three azimuth planes at elevation angles of 0$^{\circ}$ and $\pm20^{\circ}$. The plot clearly shows that the PLEs are identical using widebeam and narrowbeam antennas in either discrete angle case or over the entire azimuth plane(s). For mmWave wireless systems, many of the directional antennas (e.g., conical) or antenna arrays used have or generate very similar patterns as the horn antennas, which can be approximated by~\eqref{eq:AntPattern}. Thus the synthesizing theory of~\eqref{IREq} is validated here and can be applied to many types of antennas or antenna arrays.

\subsection{73 GHz Measurement Data}\label{73 GHz Measurement Data}
In the 28 GHz measurements, the measured elevation angles were fixed to 0$^{\circ}$ and $\pm20^{\circ}$ about the horizon, and the $\pm10^{\circ}$ elevation angles were not considered. In the measurements at 73 GHz\cite{MacCartney14:1}, however, PDPs were acquired at elevation angles separated by 5$^{\circ}$ or 8$^{\circ}$ (close to the 7$^{\circ}$ HPBW), i.e., true adjacent HPBW measurements were performed at 73 GHz. Therefore, some insight can be gained from the distribution of received power over elevation angles separated by about one HPBW using the 73 GHz measurement data.

\begin{table*}
\renewcommand{\arraystretch}{1.1}
\caption{Ratio of the received power over the strongest azimuth plane to the received power corresponding to the strongest azimuth plane plus the two adjacent azimuth planes from the 73 GHz measurements in New York City\cite{MacCartney14:1}.}
\centering
\begin{tabular}{|c|c|c|c|c|c|}
\hline
\textbf{TX Height (m)} & \textbf{RX Height (m)} & \textbf{T-R Separation (m)} & \textbf{Elevation Step ($^{\circ}$)} & \textbf{Received Power Ratio} & \textbf{Received Power Ratio in dB} \\ \hline
7 & 2 & 128 & 5 & 72.9\% & -1.4\\ \hline
7 & 2 & 139 & 5 & 76.0\% & -1.2 \\ \hline
7 & 2 & 182 & 5 & 71.9\% & -1.4\\ \hline
7 & 2 & 190 & 5 & 74.5\% & -1.3\\ \hline
7 & 4.06 & 27 & 5 & 72.0\% & -1.4\\ \hline
7 & 4.06 & 40 & 8 & 73.9\% & -1.3\\ \hline
7 & 4.06 & 74 & 5 & 72.1\% & -1.4\\ \hline
7 & 4.06 & 107 & 5 & 83.1\% & -0.8\\ \hline
7 & 4.06 & 128 & 5 & 75.3\% & -1.2\\ \hline
7 & 4.06 & 145 & 5 & 73.8\% & -1.3\\ \hline
7 & 4.06 & 182 & 5 & 73.2\% & -1.4\\ \hline
17 & 2 & 129 & 5 & 91.7\% & -0.4\\ \hline
17 & 2 & 129 & 5 & 76.7\% & -1.2\\ \hline
17 & 2 & 168 & 5 & 81.0\% & -0.9\\ \hline
17 & 4.06 & 118 & 5 & 73.9\% & -1.3\\ \hline
17 & 4.06 & 118 & 5 & 74.4\% & -1.3\\ \hline
17 & 4.06 & 127 & 5 & 91.2\% & -0.4\\ \hline
17 & 4.06 & 129 & 5 & 95.0\% & -0.2\\ \hline
17 & 4.06 & 129 & 5 & 72.8\% & -1.4\\ \hline
17 & 4.06 & 181 & 5 & 79.6\% & -1.0\\ \hline
\end{tabular}
\label{eleAngle}
\end{table*} 

Table~\ref{eleAngle} lists the ratio of the received power corresponding to the strongest azimuth plane, as compared to that corresponding to the strongest azimuth plane plus its two adjacent azimuth planes. The adjacent azimuth planes at the RX were separated by 5$^{\circ}$ or 8$^{\circ}$, close to one HPBW of the antenna. It can be observed from Table~\ref{eleAngle} that the received power over the strongest azimuth plane accounts for the majority (over 70\%) of the total received power at the strongest plus adjacent azimuth planes, with a highest ratio of 95\%, i.e., only 0.2 dB difference. Note that the adjacent azimuth planes are mostly separated by only 5$^{\circ}$ (less than one HPBW of the antenna), if the elevation angle increment increases to one HPBW, namely 7$^{\circ}$, even higher contribution is expected from the strongest azimuth plane, i.e., the ratio of the received power over the strongest azimuth plane will be even larger. Therefore, when time or resources are limited, an omnidirectional antenna pattern and resulting omnidirectional path loss may be obtained by considering the received power from directional antennas at the strongest elevation angle by scanning and summing received powers across the entire azimuth plane, or using scans across just a few adjacent elevation angles, when obtaining received powers using widebeam or narrowbeam antennas in outdoor environments. Ceiling and ground reflections  induced by indoor structures  may require more diverse elevation pointing angles to provide omnidirectional path loss data using this approach in indoor environments.

\section{Conclusion}
This paper presented and validated a method for synthesizing the omnidirectional antenna pattern and resulting omnidirectional received power and path loss from measured data using directional horn antennas, by summing the received powers from each and every measured non-overlapping directional HPBW antenna pointing angle combination. Both theoretical analyses and measured results were provided to validate this method. It was shown that the received power using 3$\times$3 narrowbeam antennas agreed relatively well with that received from a single widebeam antenna (where the azimuth and elevation HPBWs of the widebeam antenna were about three times that of the narrowbeam antenna). Using directional antennas with different beamwidths yielded almost identical received power and path loss synthesized over a systematic incremental scan of the entire azimuth plane(s). By extrapolating these results between narrowbeam and widebeam antennas for the case of a complete azimuthal scan, it is shown that the omnidirectional antenna pattern may be synthesized through azimuthal and elevation scans over the entire $4\pi$ steradian sphere. In addition, the 73 GHz measurement data showed that when considering the total received power over the strongest azimuth plane and its two adjacent azimuth planes, 72\% to 95\% of the received power arrived in the strongest azimuth plane, thus for outdoor measurements when time or resources are tight, it appears reasonable to consider the power at the strongest azimuth plane alone when comparing the received powers with different antenna beamwidths, since all but a dB or so is captured in a single azimuth plane as shown in Table~\ref{eleAngle}. The approach provided here offers substantial improvement in dynamic measurement range and coverage distance when compared to using omnidirectional low-gain antennas,  due to the ability to include antenna gains while constructing an omnidirectional antenna pattern that would otherwise have to be measured with a low-gain omnidirectional antenna with much greater transmit power.

\bibliographystyle{IEEEtran}
\bibliography{bibliography}
\end{document}